\documentclass[conference]{IEEEtran}
\IEEEoverridecommandlockouts
\usepackage{amsmath,amssymb,amsfonts}
\usepackage{algorithmic}
\usepackage{graphicx}
 \usepackage{textcomp}
\usepackage{xcolor}
\def\BibTeX{{\rm B\kern-.05em{\sc i\kern-.025em b}\kern-.08em
 T\kern-.1667em\lower.7ex\hbox{E}\kern-.125emX}}

  \makeatletter
\newcommand*{\rom}[1]{\expandafter\@slowromancap\romannumeral #1@}
\makeatother

\usepackage{xcolor}

     \usepackage[left= 2.05 cm, right= 2.05 cm, top=1.778 cm, bottom = 2.65 cm]{geometry}
\begin{document}
\bstctlcite{IEEEexample:BSTcontrol}
 %In references:@ieeetranbstctl{IEEEexample:BSTcontrol,
%  ctluse_forced_etal       = {yes},
 % ctlmax_names_forced_etal = {1},
  %ctlnames_show_etal       = {1},
  %ctldash_repeated_names   = {no}
%}
\title{Federated Learning-based MARL for Strengthening Physical-Layer Security in B5G Networks}

\author{\IEEEauthorblockN{Deemah H. Tashman\IEEEauthorrefmark{1}, Soumaya Cherkaoui\IEEEauthorrefmark{1}, and Walaa Hamouda\IEEEauthorrefmark{2}  }
\IEEEauthorblockA{\IEEEauthorrefmark{1} \small Department of Computer  and Software Engineering, Polytechnique Montreal, Montreal, Canada\\ \IEEEauthorrefmark{2}Department of  Electrical and Computer Engineering, Concordia University,  Montreal, Canada\\
Email: \IEEEauthorrefmark{1}   \{deemah.tashman, soumaya.cherkaoui\}@polymtl.ca,
\IEEEauthorrefmark{2} hamouda@ece.concordia.ca}}

\maketitle
\begin{abstract}
This paper explores the application of a federated learning-based multi-agent reinforcement learning (MARL) strategy to enhance physical-layer security (PLS) in a multi-cellular network within the context of beyond 5G networks.   At each cell, a base station (BS) operates as a deep reinforcement learning (DRL) agent that interacts with the surrounding environment to maximize the secrecy rate of legitimate users in the presence of an eavesdropper. This eavesdropper attempts to intercept the confidential information shared between the BS and its authorized users. The DRL agents are deemed to be federated since they only share their network parameters with a central server and not the private data of their legitimate users. Two DRL approaches, deep Q-network (DQN) and Reinforce deep policy gradient (RDPG), are explored and compared. The results demonstrate that RDPG converges more rapidly than DQN. In addition, we demonstrate that the proposed method outperforms the distributed DRL approach. Furthermore, the outcomes illustrate the trade-off between security and complexity.
  
\end{abstract}
\begin{IEEEkeywords}
B5G networks, 6G cellular networks, federated learning,   multi-agent deep reinforcement learning, physical-layer security.
 
\end{IEEEkeywords}
\section{Introduction}

\par\IEEEPARstart{P}{hy}sical-layer security (PLS) has been utilized in 5G as a dependable security method to assess and improve security against physical layer attacks \cite{9094381}, and it will be one of the primary components for securing 6G networks \cite{9524814,10044975}. PLS relies on channel variations to assess and enhance the privacy of sensitive data. Specifically, in the three-node wiretap model which was first proposed by Wyner \cite{9237455}, security is guaranteed as long as the channel between legitimate endpoints, i.e. the main channel, is more reliable than the one between the transmitter and malicious user (eavesdropper), referred to as the wiretap link \cite{9408651,9926102,9348134}.
%In addition to the high data rate expected in 6G networks, the reliance on AI and machine learning will be higher, which also posed additional type of security risk during training and testing \cite{9482609}.

%MARL in 6G:
In 6G networks, the reliance on machine learning (ML) will grow to be utilized to improve network performance and security. For instance,  reinforcement learning (RL) has recently been proposed as an approach for  boosting the PLS for multiple forms of wireless networks.  In RL, a single agent interacts with an environment \cite{9318243,9729992,9524882}, which is typically modeled as a Markov decision process (MDP), and discovers the optimal decision-making policy \cite{9318243,9729992}. Furthermore, deep reinforcement learning  (DRL) employs deep neural networks (DNNs) to process high-dimensional input data in order to approximate complex functions \cite{9838580,9716076,10160115}, such as the value function or policy, which are essential components of RL.
 However, future cellular networks may require the cooperation of multiple entities to avoid issues such as the non-stationary problem \cite{9738819} and increase the learning efficiency and, consequently, network performance and security. Therefore, multi-agent RL (MARL) has been proposed to address this challenge, as an agent in MARL can train quickly and benefit from the knowledge of other agents coexisting in a shared environment \cite{zhang2021multi}. In MARL, agents adapt their strategies based on the actions and behaviors of other agents.

The incorporation of ML methodologies introduces an additional category of security vulnerability throughout the training and testing phases \cite{9482609}. Specifically, sharing information between RL agents in MARL scenarios introduces a potential security vulnerability due to the possibility of unauthorized interception of this information. Recently,  federated learning (FL) has been proposed as a strategy to boost privacy when data from multiple devices needs to be aggregated \cite{9685388,10225769}. In FL,  only the training parameters are shared while the private users' data remains undisclosed \cite{mcmahan2017communication}.  In the context of FL-based DRL, agents select a specific DRL approach and subsequently communicate solely the DNN parameters to a central unit (CU). The CU collects these parameters and applies a mathematical operation, such as simple averaging. The resulting averaged DNN parameters are then distributed among the agents to update their training model. Consequently, the agents will be able to benefit from the training of other agents in a secure manner.

%{  Literature review*****:}
A limited amount of research has been conducted in the field of PLS while utilizing FL-based ML techniques. In \cite{10226140}, for instance, the authors evaluated the PLS for a wireless hierarchical FL-based supervised ML architecture in which an eavesdropper attempts to intercept data transmitted from the edge servers to the central server. The achievable data rate and the maximum uncertainty of an eavesdropper were studied. In \cite{9797923}, a wireless FL-based supervised ML was examined, and a differential privacy approach was assumed to be applied prior to aggregating users' gradients to a trusted third party (TTP). The TTP then performs secure joint source-channel coding to prevent eavesdropping before transmitting the codeword to a central server. An eavesdropper was presumed in this work  to wiretap the received signal from the central server. The authors demonstrated that security is achieved as the length of the codeword reaches infinity.  
To the greatest extent of our understanding, only these two research papers have considered the integration of PLS and FL-based ML. 
 %Recently, in \cite{9677132} has studied a FL-based RL to improve the sum rate of a cellular network. However, the security of the mobile users at each cell against eavesdropping attack was not considered.

%Our contribution:
To the authors' knowledge, no previous research has evaluated the security of an FL-based DRL for cellular networks. Therefore, we propose an FL-based MARL for a multi-cellular network. At each cell, a base station (BS) serves multiple legitimate mobile users in the presence of a passive eavesdropper attempting to intercept the data exchanged between the BS and legitimate users. Each BS functions as a DRL agent with the potential of maximizing the secrecy sum rate by controlling its transmission power. To safeguard the users' privacy, these agents share their DNN parameters with the central server unit without sharing users' information. The CU then aggregates and averages these parameters. The averaged parameter is then communicated to the DRL agents to adjust their training.  Therefore, these agents profit from the knowledge of other agents in order to learn faster and more securely. 
Our motivation stems from the belief that the FL strategy targets preserving the security of users' data shared between the BSs and the central server; however, the security of data shared between the BSs and legitimate users remains vulnerable to eavesdropping. To maximize the secrecy sum rate of the cellular network, we use two DRL techniques: the deep Q-network (DQN) and the Reinforce deep policy gradient (RDPG).

The remaining sections of the paper are as follows; the system model is described in section \rom{2}. In section \rom{3}, the maximization of the secrecy sum rate problem is formulated. The DRL solution based on FL is presented in section \rom{4}. The numerical results are presented in section \rom{5}, while the conclusions are presented in section \rom{6}.

\section{System Model}
\begin{figure}[t]
  \centering
\includegraphics[width=0.5\textwidth]{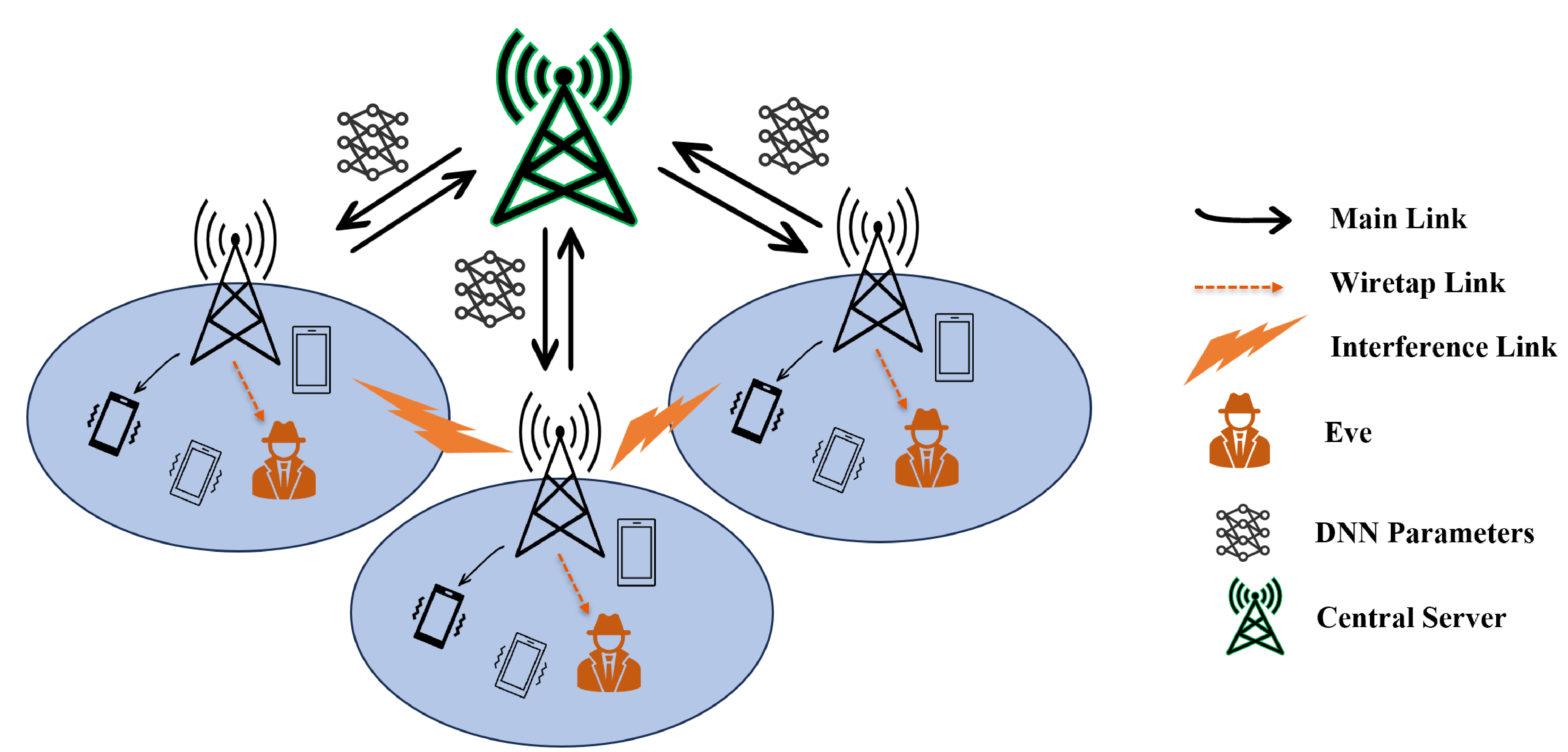}
  \caption{The  system model.}
  \label{sys}
\end{figure}

As depicted in Fig. \ref{sys}, a cellular network consists of $B$ $(b \in 1,2,\cdots,B)$ base stations (BSs), each of which serves $L$ $(l \in 1,2,\cdots,L)$ mobile users. At each BS, there is an eavesdropper (Eve) attempting to wiretap the confidential information communicated between the BS and each of the authorized users.  The channel  gain between each BS $(b)$ and legitimate user $(l)$ at time slot $t$ is denoted by
\begin{eqnarray}
g_{b,l,c}^t=\left|h_{b,l,c}^t\right|^2 {\rho}_{b,l,c},
\end{eqnarray}
\noindent where  $c$ denotes the index of a specific cell. The term $\left|h_{b,l,c}^t\right|^2$ indicates the power gain of the channel, while ${\rho}_{b,l,c}$ denotes the large-scale fading element. This element encompasses both the path loss and the log-normal shadowing effects. In a similar vein, the wiretap channel gain, namely the one that exists between BS $(b)$ and eavesdropper (Eve) at cell $c$ and time slot $t$, is expressed as
\begin{eqnarray}
g_{b,Eve,c}^t=\left|h_{b,Eve,c}^t\right|^2 {\rho}_{b,Eve,c},
\end{eqnarray}
\noindent where $\left|h_{b,Eve,c}^t\right|^2$ denotes the wiretap channel power gain and ${\rho}_{b,Eve,c}$ represents the large-scale fading, which includes the path loss and log-normal shadowing. 
The small-scale fading for the  main and wiretap links are both exhibiting Rayleigh fading distribution and thus the probability distribution function (PDF) of the channel power gain   follows the exponential distribution  as 
\begin{eqnarray}  
f_{\left|h_{b,i,c}^t\right|^2} (y)=   \lambda_{i}  \exp\left(-\lambda_{i} y\right),
\end{eqnarray}
\noindent for $i \in \{l,Eve\}$ and $\lambda_{i}$ is the corresponding fading channel parameter. 
The main link capacity, i.e. the data rate at user $l$, is given by
\begin{eqnarray}  
R_{l,b}=   \log_2\left(1+\gamma_{l,b}^t\right),
\end{eqnarray}
\noindent where  $\gamma_{l,b}^t$ denotes the signal-to-interference-plus-noise-ratio (SINR) of user $l$ in cell $b$ during time slot $t$. The SINR is mathematically stated as
\begin{eqnarray}  
\gamma_{l,b}^t= \frac{p_{b,l}^t g_{b,l,c}^t}{I^t+N_0},
\end{eqnarray}
\noindent where $p_{b,l}^t$ signifies the transmission power from base station $b$ to user $l$ during time slot $t$.   $N_0$ represents the variance of the additive-white-Gaussian noise (AWGN) with zero mean. Additionally, $I^t$ represents the combined inter-cell and intra-cell interference and it is expressed as \cite{9677132}
\begin{eqnarray} \label{intersum} 
I^t= \sum_{i=1,i\neq l}^L p_{b,i}^t g_{b,i,b}^t +\sum_{i=1,i \neq b}^B \sum_{c} p_{i,c}^t g_{i,c,l}^t .
\end{eqnarray}
\noindent The first component of equation (\ref{intersum}) indicates the intra-cell interference, which arises from the simultaneous transmission of several users inside the same cell. Moreover, the second component of the equation represents inter-cell interference, which occurs when a user in one cell experiences interference from neighboring cells.
The wiretap channel capacity is given as
\begin{eqnarray}  
R_{Eve,b}=   \log_2\left(1+\gamma_{Eve,b}^t\right),
\end{eqnarray}

\noindent where $\gamma_{Eve,b}^t$ is the SINR of the eavesdropper in cell $b$.  For analyzing the system under the worst-case scenario, it is presumed that only noise impairs Eve's reception quality. Thus,  $\gamma_{Eve,b}^t$   is given by
\begin{eqnarray}  
\gamma_{Eve,b}^t= \frac{p_{b,l}^t g_{b,Eve,c}^t}{N_0}.
\end{eqnarray}
\noindent
%with  $p_{b,Eve}^t$ being the transmission power from base station $b$ to the eavesdropper during time slot $t$.
\section{Maximizing Secrecy Rate}

Given the presence of an eavesdropper at each cell, the primary goal of each agent (BS) is to optimize the privacy of the authorized users by determining an optimal allocation of transmission power from each BS to the respective users. The PLS is utilized  to evaluate the level of privacy associated with the information exchanged between mobile users and their respective BS.  Hence, the problem is formulated as
 \begin{align} \label{opti-prob}
   \mathcal{P}: \;\;  & \underset{p_{b,l}^t}{\text{max}}
    & & \sum_{b} \sum_{l} C_{b,l}^{sec} \\
    & \text{s.t.}  
    && \label{firstrate1} 0 \leq p_{b,l}^t \leq P^{max}, & 
  \end{align}
\noindent where $P^{max}$ represents the maximum amount of transmission power of the  BS. The variable $C_{b,l}^{sec}$ represents the secrecy capacity, which refers to the highest achievable transmission rate for confidential data via a communication channel while ensuring protection against possible eavesdroppers \cite{9612017,9500621,9838746,10278964}. The expression for $C_{b,l}^{sec}$ is provided as \cite{10182973}
\begin{eqnarray} \label{cap-eq}
C_{b,l}^{sec}= \max \{R_{l,b}-R_{Eve,b},0\}.
\end{eqnarray} 

\section{Federated Learning-based DRL Solution}
Given the non-convex and complex nature of the problem stated in (\ref{opti-prob}), it becomes necessary to reframe it as a 
MARL problem. In this formulation, each BS operates as an agent employing DRL techniques. Every DRL agent aims to maximize the security of the users it serves in the presence of Eve, while also taking into account the potential interference that may affect nearby cells.  In order to accomplish this objective, the DRL agent will allocate distinct transmission power levels to its users throughout each time slot.   
In this section, we use two methodologies of DRL to address our optimization problem,  namely DQN and RDPG.

The problem presented in  (\ref{opti-prob}) may be formulated as a model-free MDP. This is due to the fact that the state of each slot is dependent only on the state of the previous slot, thereby fulfilling the Markov property \cite{10182973}. Hence, the model encompasses the four tuples denoted as $<S, A, R, T>$. The representation of the space of states is denoted as $S$, whereas the state at each time slot is indicated as $S_t$. The available state space encompasses the channel gains between the BSs and users, as well as the prior transmission power and secrecy rate as 
\begin{eqnarray} \label{states}
S_t=  \{g_{b,i,c}^t,p_{b,l}^{t-1},C_{b,l}^{t-1}\}.
\end{eqnarray} 

\noindent For $i\in\{l,Eve\}$. Furthermore, the symbol $A$ denotes the action space, which encompasses the set of feasible actions available for the agent to choose from. In each time slot, the agent is required to make a decision about the transmission power level for each of its users. This decision is made by choosing a value from the action space, which consists of a set of $\eta$ power levels. Additionally, it is assumed that all agents possess an identical action space.  The variable $A_t$ denotes the action taken at each time interval and is mathematically represented as
\begin{IEEEeqnarray}{lcr}
   A_t=\{0,\frac{P_{max}}{\eta-1},\frac{2P_{max}}{\eta-1},\cdots,P_{max}\}.
\end{IEEEeqnarray}

\noindent The reward function at each time slot is expressed as
\begin{IEEEeqnarray}{lcr}
   R_t=\sum_{l} {C_{b,l}^{sec,t}}+  \sum_{b'}  \sum_{l}R_{l,b'}^t,
\end{IEEEeqnarray}

\noindent where  $\sum_{b'} \sum_{l} R_{l,b'}^t$ is used to consider the  interference impact on the adjacent cells.   To maximize the  cumulative reward, the agent's policy develops to implicitly achieve a balance between maximizing its secrecy rate and minimizing the interference produced to neighboring cells.
Time step $T$ refers to a set of discrete time intervals. The process of moving from state $s_t$ to the subsequent state $(s_{t+1})$ is referred to as a step. The state-action combination is iteratively executed until all time slots have elapsed.

To accelerate the learning process, we also employ the FL method, in which all agents periodically share their trained NN parameters with a central server. Moreover, the FL approach assures that these federated agents (BSs) share only their NN parameters and not their users' private data. This will guarantee the confidentiality of the shared private conversations between the BSs and the central server.  In this section, we describe the FL-based DQN and FL-based RDPG methodologies to safeguard the transmission netween the BS and users.

\subsection{Federated Learning-based DQN}
The primary objective of our  DQN strategy is to optimize the training of the agent by facilitating its interaction with the environment, to maximize the cumulative reward in the long term \cite{9351818,9729992}. A DNN is used to calculate the estimated Q-values for every possible state-action in the environment, to approximate the optimum Q-function \cite{9524882}. The learning process involves the acquisition of knowledge on how to associate state-action pairings with Q-values, which serve as estimates of the anticipated reward for executing a certain action in a given state. This estimation takes into consideration the potential future rewards associated with adhering to a particular policy.  In DQN, the Q function is given by \cite{9645987}
\begin{IEEEeqnarray}{lCr}  \label{Q-ini}
  Q^{\pi}\left(s,a\right)= E_{s^{\prime}, a^{\prime}}\left( r+\gamma Q^{\pi} \left(s^{\prime},a^{\prime}\right)|s,a\right),
\end{IEEEeqnarray}
\noindent where $s$ and $a$ represent the present state and the chosen action, respectively. $a^{\prime}$ is the action chosen at state $s^\prime$, which is the state reached after the action $a$ is performed. $r$ symbolizes the immediate reward for choosing action $a$. $E[\cdot]$ is the expectation operator. $\gamma$ is  the discount factor, which is a scalar value between 0 and 1 that dictates the significance of future rewards. 
 The discount factor guarantees that the value function converges and that the agent always receives a finite value, even when the number of time steps approaches infinity \cite{andrew1999reinforcement}. $\pi$ represents the agent's strategy for selecting actions in a specific state. The policy determines which action the agent should perform in each state to maximize the expected reward. Due to the extremely dynamic environment, the $\epsilon$-greedy strategy is used to determine the optimal transmit power of the BS. During training, the algorithm employs an $\epsilon$-greedy exploration policy, where the agent chooses a random action with probability $\epsilon$ and the action that maximizes the Q-value for the current state with probability $(1-\epsilon)$.

In order to optimize the overall long-term return, it is necessary to identify the most advantageous course of action for each time interval, with the objective of maximizing the state-action value described in (\ref{Q-ini}) as
\begin{IEEEeqnarray}{lCr}  
  a^\ast =\underset{a} \arg \max  Q^{\pi}\left(s,a\right) .
\end{IEEEeqnarray}
\noindent In DQN algorithm, DNNs are used to estimate the Q-value function. This is achieved by iteratively updating the parameter $\theta_D$, which corresponds to the weights and biases of the NN.  The approximation is expressed as 
\begin{IEEEeqnarray}{lCr}  
   Q\left(s,a;\theta_D \right)\approx  Q^{\ast}\left(s,a\right) ,
\end{IEEEeqnarray}
\noindent where $ Q^{\ast}\left(s,a\right)=E_{s^{\prime}} \left(r+\gamma \max_{a^{\prime}} Q^{\ast}\left(s^{\prime},a^{\prime}|s,a\right)\right)$ denotes the optimal value function. 
The parameter $\theta_D$ is iteratively adjusted using the gradient descent optimization algorithm and the backpropagation technique. This adjustment is based on the loss between the predicted Q-values and the desired Q-values. The loss function denoted as $\mathcal{L}$ and defined as the mean square error is given by
\begin{IEEEeqnarray}{lCr}  \label{loss}
  \mathcal{L(\theta_D)}=E(( r_s+\gamma \max_{a^{\prime}} Q^{\pi} \left(s^{\prime},a^{\prime};\theta_D \right)- \nonumber \\ Q^{\pi}\left(s,a;\theta_D\right)
   )^2),    
\end{IEEEeqnarray}
\noindent with the first portion of (\ref{loss})  $\left( r_s+\gamma \max_{a^{\prime}} Q^{\pi} \left(s^{\prime},a^{\prime};\theta_D \right)\right)$ denoting the target Q-value, while the second component is the predicted Q-value, which is updated periodically throughout the learning process. $r_s$ signifies the reward received for transitioning from state $s$ to state $s^\prime$.
The loss function can be minimized via stochastic gradient descent (SGD), which adjusts the value of $\theta_D$ in the direction that minimizes the loss.
 The SGD rule is given as 
 \begin{IEEEeqnarray}{lCr}  \label{sgd}
y=\alpha \frac{\partial  \mathcal{L(\theta_D)} }{\partial \theta_D}    ,
\end{IEEEeqnarray}
\noindent where $0<\alpha<1$  depicts the learning rate, which dictates the rate at which the Q-value function is updated relying on the observed rewards and the present estimates of the Q-value function. Finally, using (\ref{sgd}), the value of $\theta_D$ is updated as   
\begin{IEEEeqnarray}{lCr}   
\theta_D=\theta_D-y .
\end{IEEEeqnarray}

The FL objective function   essentially seeks to find the local model weights  that minimize the overall loss across all BSs. It takes into account both the local loss function and its DNN weights. Hence, the global loss function is given by \cite{9797923}

 \begin{IEEEeqnarray}{lCr} \label{cost1}  
\min_{\theta_D} \mathcal{L}(\theta_D)=\sum_{i=1}^B \nu_i \mathcal{L}_i (\theta_{Di}),
\end{IEEEeqnarray}
\noindent where $\nu_i$ represents the contribution of cell $i$ to the network, such that $\nu_i=\frac{l_i}{\sum_{i=1}^B l_i}$, with $l_i$ being the number of users in cell $i$.

%After a certain number of episodes, agents will share their DNN parameters  with a central server in order to fasten the training process. We denote this parameter by $\xi$. These agents are federated in the sense that only the DNN parameters will be shared and not the users' private data.  At the central server, these DNN parameters will be averaged and the result of this process will be sent back to all the agents so they can update their models accordingly. This way all agents will be benefiting from the training of other agents without sharing the users' confidential information,

 \subsection{Federated Learning-based RDPG}

Reinforce is a DRL technique that trains a policy (strategy) by evaluating its performance in real-world situations in relation to the rewards it receives.   Reinforce functions with a stochastic policy that generates probabilistic decisions in each state, encouraging exploration and adaptability. It functions by having the agent's NN generate a probability distribution over possible actions, from which the agent samples its action. Reinforce is a flexible and exploratory decision-making method as the distribution's probabilities are refined during training to maximize expected rewards. In other words, if a policy produces favorable results (high rewards), it is reinforced by adjusting its parameters (DNN weights) to continue those actions. If, on the other hand, the policy results in unfavorable outcomes (low rewards), it is discouraged and its parameters are altered to prevent those actions \cite{9120241}.  
The objective of the RDPG is to reduce the loss function by taking the gradient with respect to the DNN parameter $(\theta_R)$  as
\begin{IEEEeqnarray}{lCr}  \label{loss2}
 \triangledown_{\theta_R} \mathcal{Y(\theta_R)}=E\{\triangledown_{\theta_R} \log\left(\pi(a|s;\theta_R)R\right)\} ,
\end{IEEEeqnarray}
\noindent where $\pi(a,|s;\theta_R)$ is the parameterized policy \cite{9677132}.
Similar to (\ref{cost1}), the RDPG cost function is given as 
 \begin{IEEEeqnarray}{lCr} \label{cost2}  
\min_{\theta_R} \mathcal{Y}(\theta_R)=\sum_{i=1}^B \nu_i \mathcal{Y}_i (\theta_{Ri}).
\end{IEEEeqnarray}

%------------------------------------------------
 
 \section{Numerical Results} 
\par This section provides the simulation for our proposed system model and analyses. The parameters considered in this section, unless otherwise specified in the figures' caption are: $B=25$,  $L=4$, $\xi=100$, $P^{max}=38$ dBm, $\eta=10$, $N_0=1$, $\alpha=0.001$, $\gamma=0.99$, and $\lambda_i=1.5$

\begin{figure}   [b]
  \centering
  \includegraphics[width=1.0\linewidth]{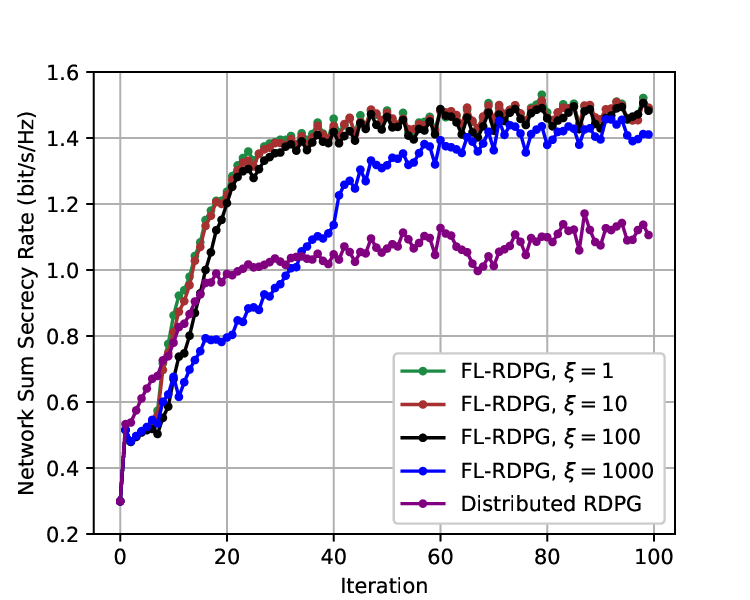}
  \caption{Network sum secrecy rate for FL-RDPG and distributed RDPG for different values of $\xi$.}
  \label{agg}
\end{figure}

The comparison between the FL-based RDPG (FL-RDPG) and the distributed RDPG is depicted in Fig. \ref{agg}. Distributed RDPG implies that each agent executes the RDPG independently, without exchanging data with a central server. One can conclude that the FL-based RDPG outperforms the distributed RL since agents (BSs) cooperate by sharing their DNN parameters in the FL approach. When agents receive updated DNN parameters from the CU, they periodically share their acquired knowledge in a manner that protects their privacy, resulting in enhanced security.  Distributed RL agents, in contrast, operate in isolation and cannot utilize the knowledge gained by other agents. In addition, when the aggregation parameter $(\xi)$ increases, indicating that agents share their parameters less frequently, security deteriorates and agents reach convergence slowly. This is because lower cooperation attempts with the other agents hinders the exchange of vital information required for secure and effective learning.

\begin{figure}   [t]
  \centering
  \includegraphics[width=1.0\linewidth]{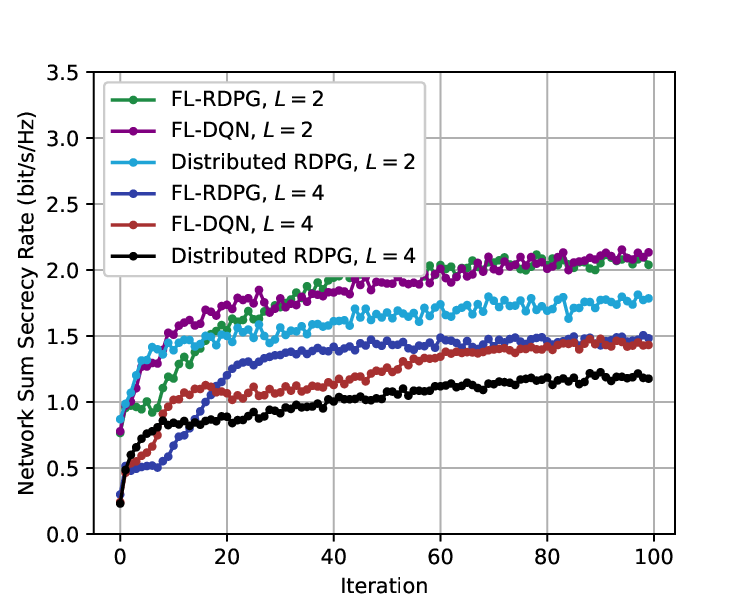}
  \caption{Network sum secrecy rate for FL-RDPG, FL-DQN and distributed RDPG for different values of $L$.}
  \label{L}
\end{figure}

 Fig. \ref{L} compares FL-RDPG and FL-based DQN (FL-DQN) for various values of the number of legitimate users in each cell $(L)$. It is evident that FL-RDPG converges faster than FL-DQN. This is because DQN employs the $\epsilon$-greedy policy, which compels the agent to completely explore the environment at the start of training to discover all possible rewards for its actions at each state. However, the agent in the RDPG approach does not utilize this method, as it is more sample-efficient and encourages exploration more organically. 
In situations where eavesdroppers are present and attempting to intercept communication, rapid convergence is of the utmost importance. A more swiftly converging algorithm can modify its rules more quickly, thereby increasing the efficiency of secrecy measures. FL-RDPG demonstrates enhanced capabilities in responding to eavesdropping efforts, thereby making it more challenging for potential eavesdroppers to decode sensitive data. Moreover, an increase in the value of  $L$ leads to a degradation in security due to the concurrent rise in intra-cell and inter-cell interference. Increased interference negatively impacts the reliability of the links used by authorized users, hence diminishing their privacy.

\begin{figure}  [b] 
  \centering
  \includegraphics[width=1.0\linewidth]{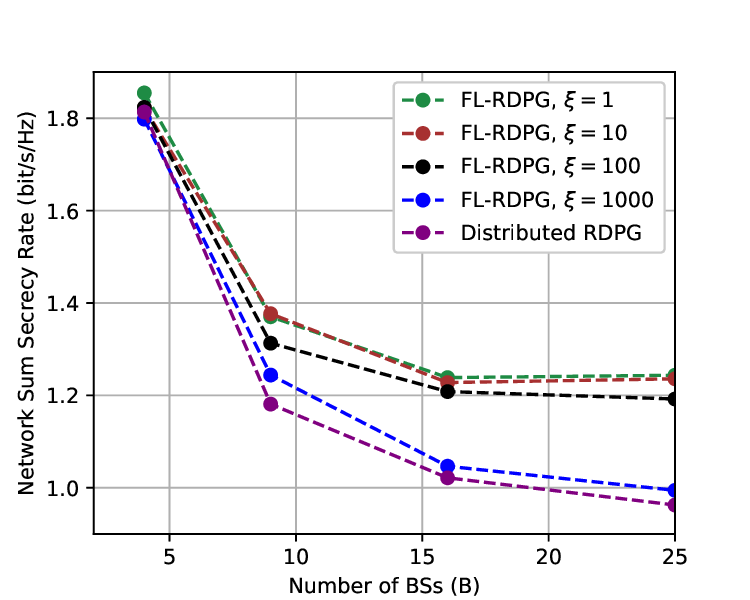}
  \caption{Network sum secrecy rate for FL-RDPG  distributed RDPG versus the number of BSs $(B)$.}
  \label{aggN}
\end{figure}

Fig. \ref{aggN} illustrates the influence of the number of base stations $(B)$ on the security. The level of privacy for shared information is shown to degrade with an increase in the parameter $B$. This phenomenon is due to the direct relationship between the rise in   $B$ and the subsequent increase in the number of nearby cells. Consequently, the interference experienced by these surrounding cells is also increased. This will compromise the reliability of the connection of authorized users, thus diminishing their level of privacy. Furthermore, the alteration of the variable $\xi$ in the context of a small value of $B$ (number of agents) has a negligible effect on security. Nevertheless, when the value of $B$ grows, the parameter $\xi$ begins to significantly affect the security. This implies that in scenarios with a limited number of agents, namely when the value of $B$ is low, increasing the frequency of aggregation (by decreasing $\xi$) does not significantly affect security. In this scenario, it is recommended to reduce the aggregation frequency (by increasing the value of $\xi$) to minimize complexity. Nevertheless, in the case of a substantial number of agents ($B$), aggregating more frequently is suggested.  This modification has the potential to optimize network performance and boost security, particularly in complex multi-agent scenarios. Hence, the trade-off between security and complexity is reliant on the scale of the network $(B)$.

\section{Conclusions}
In this paper,  PLS is utilized to evaluate the privacy of a multi-cellular network employing an FL-based MARL methodology.
We compared the FL-based DQN, the FL-based RDPG, and the distributed DRL. We demonstrated that the FL-based approaches outperform the distributed DRL, indicating that cooperation between agents is encouraged to improve the privacy of users in the presence of an eavesdropper. We also observed that the FL-based RDPG converges faster than the FL-based DQN.  In addition, as the aggregation parameter increases, security degrades due to agents' delayed learning as a result of fewer cooperative attempts.  We also drew the conclusion that security degrades as the number of legitimate users or the number of cells increases, as interference increases. Finally, we observed that the aggregation frequency is dependent on the number of cells (BSs), as it is recommended to aggregate the DNN parameters less frequently if the number of cells is small to reduce the network's complexity, illustrating the trade-off between security and complexity.


\end{document}